\documentclass[10pt,onecolumn,aps,prd,preprintnumbers,showpacs,superscriptaddress,nofootinbib,amsmath,amssymb,floats,floatfix,showkeys,notitlepage,longbibliography]{revtex4-2} 
\usepackage{orcidlink}
\usepackage{comment}
\usepackage{lipsum}
\usepackage{graphicx}
\usepackage{subfigure}
\usepackage{palatino}
\usepackage{sans}
\usepackage{hyperref}
\hypersetup{colorlinks=true,linkcolor=blue,urlcolor=blue,citecolor=blue}
\usepackage[toc,page]{appendix}
\usepackage[normalem]{ulem}
\usepackage{adjustbox}
\usepackage{latexsym}
\usepackage{amsmath}
\numberwithin{equation}{section}
\usepackage{amssymb}
\usepackage{amsfonts}
\usepackage{dcolumn}
\usepackage{bm}
\usepackage{tikz}
\usetikzlibrary{decorations.pathmorphing}
\usepackage{bigints}
\usepackage{array,tabularx,multirow,booktabs}
\usepackage[tracking=true]{microtype}
\usepackage{soul} 
\SetTracking{}{500}
\SetTracking{encoding={*}, shape=sc}{40}
\UseRawInputEncoding 
\allowdisplaybreaks
\usepackage[utf8]{inputenc}
\usepackage{xcolor} 

\begin{document} \sloppy

\title{Radial Integral Reformulation of the Gauss-Bonnet Weak Deflection Angle at Finite Distance}

\author{Ali \"Ovg\"un \orcidlink{0000-0002-9889-342X}}
\email{ali.ovgun@emu.edu.tr}
\affiliation{Physics Department, Eastern Mediterranean University, Famagusta, 99628 North
Cyprus via Mersin 10, Turkiye.}

\author{Reggie C. Pantig \orcidlink{0000-0002-3101-8591}} 
\email{rcpantig@mapua.edu.ph}
\affiliation{Physics Department, School of Foundational Studies and Education, Map\'ua University, 658 Muralla St., Intramuros, Manila 1002, Philippines.}

\begin{abstract}
We develop a radial integral reformulation of finite distance gravitational lensing in optical geometry for static, spherically symmetric spacetimes. Starting from the Gauss-Bonnet characterization of the finite distance deflection angle, we adopt the Li-type curvature primitive identity [https://doi.org/10.1103/PhysRevD.101.124058] [2006.13047], which reduces the curvature-area contribution to a one-dimensional integral evaluated along the physical light ray. We then remove the remaining implicit orbit dependence by an explicit change of variables using the null first integrals, converting the Li line integral from $\phi$-integration to $r$-integration and splitting the trajectory at the turning point (closest approach). The resulting formula expresses the deflection angle as a sum of two radial integrals over $[r_0,r_S]$ and $[r_0,r_R]$ plus the finite distance angular bookkeeping term, with a transparent normalization/cancellation structure for the curvature primitive. In Schwarzschild gauge, we provide a weak-field evaluation toolkit that reduces the computation to reusable families of standard radial integrals and gives compact expressions for the endpoint incidence angles in the optical metric. Worked examples include Schwarzschild and Kottler (Schwarzschild-de Sitter) spacetimes and a black hole immersed in perfect-fluid dark matter with a finite halo. For the finite-halo model we derive closed-form leading weak-deflection expressions for mixed endpoint configurations (source/receiver inside or outside the halo), illustrating the modularity of the radial Gauss-Bonnet pipeline. 
\end{abstract}


\maketitle

\section{Introduction} \label{sec1}

Gravitational lensing is one of the most robust probes of spacetime geometry: in the weak-field regime the deflection of light encodes the integrated influence of curvature along null propagation, while in strong-field regimes it reveals the global structure of black-hole photon regions and the onset of multiple imaging \cite{Bozza:2002zj,Bozza:2010xqn,Virbhadra:1999nm,Virbhadra:2008ws,Claudel:2000yi}.
Traditional derivations of the bending angle proceed by solving the null geodesic equation and comparing the asymptotic directions of the ray at past and future infinity, as in standard treatments of ray optics and general relativity \cite{perlick2003ray,Wald:1984rg}.
Although this approach is powerful in asymptotically flat spacetimes, it is not always well matched to realistic observation: many configurations involve emitters and receivers at finite radii, and a wide class of physically relevant models (finite halos, nontrivial asymptotics, or cosmological patches with horizons) do not admit a clean \emph{infinity-to-infinity} comparison \cite{Ishihara:2016vdc,Ono:2019hkw,Rindler:2007zz,Ovgun:2025mdg,Lu:2025mcm}.

Optical geometry provides an appealing alternative viewpoint. For static spacetimes, null rays project to geodesics of a two-dimensional Riemannian optical metric on the spatial section \cite{perlick2003ray,Gibbons:2008rj}.
In stationary settings, the corresponding optical description is naturally formulated in Randers/Zermelo data (i.e.\ a Finslerian structure), which enables a geometric lensing treatment beyond purely static metrics \cite{Gibbons:2008zi,Werner:2012rc}.
In this setting, the bending angle can be formulated intrinsically in terms of local angles measured at the endpoints and the geometric data of the optical manifold \cite{Ishihara:2016vdc,Arakida:2017hrm}.
A particularly elegant development is the use of the Gauss-Bonnet theorem, which relates an integrated curvature over a surface to boundary terms and corner contributions \cite{Gibbons:2008rj,Chern_1944,Allendoerfer_1943}.
This converts lensing into a global geometric statement: the deflection angle is determined by the Gaussian curvature of the optical metric integrated over an appropriately chosen domain, together with boundary geodesic curvature and jump angles \cite{Gibbons:2008rj,Arakida:2017hrm}.

Finite distance lensing can be formulated cleanly within this framework by placing the source $S$ and receiver $R$ at finite radii and defining the deflection angle through the optical incidence angles at $S$ and $R$ plus the coordinate separation in $\phi$ \cite{Ishihara:2016vdc,Ono:2019hkw}.
This definition is coordinate-invariant once the angles are computed in the optical geometry, and it naturally tracks finite distance corrections that are otherwise hidden in asymptotic constructions \cite{Ishihara:2016vdc,Ishihara:2016sfv,Ono:2017pie}.
It is also well suited for spacetimes where the interpretation of bending at infinity is subtle, such as Kottler-type backgrounds where local angle definitions play an essential role \cite{Rindler:2007zz,Adler:2022qtb,Khriplovich:2008ij,Park:2008ih,Sereno:2007rm,Arakida:2011ty,Ishak:2008zc,Bhadra:2010jr,Sereno:2008kk}.
The cost, however, is computational: the Gauss-Bonnet area term typically requires an explicit parametrization of the surface bounded by the ray and auxiliary curves, and this can become cumbersome even in weak-field expansions \cite{Gibbons:2008rj,Ishihara:2016vdc,Jia:2020xbc}.

A major step toward practical implementation is the Li-type curvature primitive identity \cite{Li:2020wvn}.
For a broad class of two-dimensional optical metrics relevant to static, spherically symmetric spacetimes, the combination $\mathcal K \sqrt{\det g}$ (Gaussian curvature times the area density) can be written as an exact radial derivative, reducing the curvature-area term to a one-dimensional integral of a radial primitive evaluated \emph{along the physical ray} \cite{Li:2020wvn}.
This reduction is complementary to coordinate-based streamlining strategies, such as Cartesian-like implementations of Werner's method, which can simplify intermediate steps in practical computations \cite{Li:2024ujw}.
Conceptually, these developments sharpen the relationship between optical-geometry lensing and more traditional geodesic approaches, and equivalence results provide useful extension on derived deflection formulas \cite{Li:2019mqw,Li:2020wvn}.
Nonetheless, a residual obstruction remains. The Li-reduced line integral is naturally written as $\int \mathcal P(r(\phi))\,d\phi$, and therefore still depends on the orbit shape in the implicit form $r(\phi)$.
In practice one is often forced back into orbit inversion, iterative substitutions, or repeated perturbative reconstructions of $r(\phi)$, partly defeating the original purpose of the reduction \cite{Jia:2020xbc,Li:2019mqw,Li:2020wvn}.

The present work addresses this bottleneck. We provide a complete derivation of a \emph{radial integral reformulation} of finite distance optical-geometry lensing that removes the remaining orbit dependence at the integrand level.
The essential idea is straightforward but, to our knowledge, has not been implemented in a systematic Gauss-Bonnet pipeline: since the orbit is known in differential form through null first integrals, one may convert $d\phi$ to $(d\phi/dr)\,dr$ along each monotone branch of the trajectory.
The only subtlety is the turning point at the closest approach radius $r=r_0$, where $dr/d\lambda$ changes sign.
By splitting the ray into incoming and outgoing branches and using an absolute orbit factor $|d\phi/dr|$, we obtain a representation in which the Li term is expressed as a sum of two radial integrals, one over $[r_0,r_S]$ and one over $[r_0,r_R]$ \cite{Li:2020wvn}.
This produces a modular endpoint decomposition and yields a natural toolkit for weak-field evaluation: once $\mathcal P_{\rm N}(r)$ (a normalized primitive) and $|d\phi/dr|$ are expanded, the deflection reduces to a small set of standard integral families \cite{Jia:2020xbc}.

Beyond computational efficiency, the radial formulation clarifies the cancellation structure inherent in curvature-based lensing.
The curvature primitive is defined only up to an additive constant; without careful normalization, spurious background contributions can appear.
We make the normalization explicit and show how it enforces integrand-level cancellations against the kinematic $\phi_{RS}$ term in a perturbatively consistent manner.
This becomes especially valuable for piecewise geometries, such as finite halos, where the metric (and hence the primitive) changes form across a matching radius.
In that setting the radial representation makes the necessary splitting and endpoint taxonomy transparent.

Finally, we emphasize that the optical-geometry viewpoint is not restricted to vacuum light propagation.
Unified Gauss-Bonnet treatments exist for lensing in dispersive media (e.g.\ plasma) and for massive-particle deflection, where optical/Jacobi geometrizations provide a natural language for endpoint-sensitive observables \cite{Bisnovatyi-Kogan:2010flt,Crisnejo:2018uyn,Gibbons:2015qja,Li:2019qyb}.
Moreover, lensing/shadow diagnostics have been widely applied across modified gravity models and exotic compact objects, underscoring the utility of efficient, transferable deflection pipelines \cite{Kumar:2020hgm,Fu:2021akc,Nascimento:2020ime,Kuang:2022ojj,Sarkar:2025fao,Capozziello:2025wwl,DeBianchi:2025bgn,Capozziello:2024ucm}.

The paper is organized as follows.
Section \ref{sec2} reviews finite distance lensing in optical geometry and the construction of the optical metric on the equatorial plane.
Section \ref{sec3} presents the radial reformulation with a complete derivation, starting from null first integrals and culminating in the turning-point split formula, together with a discussion of primitive normalization and a specialization to Schwarzschild gauge.
Section \ref{sec4} develops a weak-field evaluation toolkit that reduces the problem to standard radial integral families and gives explicit endpoint-angle relations.
Section \ref{sec5} provides worked examples, including Schwarzschild and Kottler spacetimes and a finite-halo PFDM model with mixed endpoint configurations.
Section \ref{sec6} concludes with a discussion of scope, limitations, and extensions.

\section{Finite distance lensing in optical geometry} \label{sec2}
This section summarizes the optical-geometry framework for finite distance weak lensing and sets the notation needed for the curvature primitives used later.

\subsection{Optical metric on the equatorial plane} \label{sec2.1}
We consider a static, spherically symmetric spacetime and restrict to the equatorial plane
($\theta=\pi/2$), so that the relevant $(t,r,\phi)$ sector can be written as
\begin{equation} \label{2.1}
ds^{2}=-A(r)\,dt^{2}+B(r)\,dr^{2}+C(r)\,d\phi^{2},
\qquad A(r)>0 \ \text{in the static region}.
\end{equation}
For null propagation, $ds^{2}=0$, and the coordinate time $t$ plays the role of an optical
length along the ray. Solving for $dt^{2}$ yields the (two-dimensional) optical metric on the
spatial section,
\begin{equation} \label{2.2}
dt^{2}\equiv dl^{2}=g^{(\mathrm{opt})}_{ij}dx^{i}dx^{j}
=\frac{B(r)}{A(r)}\,dr^{2}+\frac{C(r)}{A(r)}\,d\phi^{2}.
\end{equation}
This Riemannian metric governs the spatial light path as a geodesic of
$(\mathcal M,g^{(\mathrm{opt})}_{ij})$. Explicitly, the nonzero components are
$g_{rr}=B/A$ and $g_{\phi\phi}=C/A$, so that
\begin{equation} \label{2.3}
\det g=\frac{B(r)C(r)}{A(r)^{2}}, \qquad
dS=\sqrt{\det g}\,dr\,d\phi=\frac{\sqrt{B(r)C(r)}}{A(r)}\,dr\,d\phi .
\end{equation}
Because the metric functions depend only on $r$, the Levi-Civita connection is fixed by radial
derivatives; in particular,
$\Gamma^{\phi}{}_{r\phi}=\tfrac12\,\partial_{r}\ln\!\bigl(C/A\bigr)$, which will enter the
curvature primitive introduced later. This setup is agnostic to asymptotic flatness: finite distance
lensing is formulated by placing the source and receiver at radii $r_S$ and $r_R$ within the
admissible domain $A(r)>0$, and working intrinsically on the optical manifold.

\subsection{Deflection angle at finite distance} \label{sec2.2}
Following the finite distance definition of Ishihara-Suzuki-Ono-Asada, the total deflection angle is defined geometrically by \cite{Ishihara:2016vdc}
\begin{equation} \label{2.4}
\alpha \equiv \Psi_R-\Psi_S+\phi_{RS}, \qquad \phi_{RS}\equiv \phi_R-\phi_S,
\end{equation}
where $\Psi_S$ ($\Psi_R$) is the angle between the tangent to the light ray and the outward radial direction at the source (receiver), both measured in the optical manifold.

To connect $\alpha$ to curvature, we apply the Gauss-Bonnet theorem to a compact, simply connected domain $D$ in the optical geometry. A convenient choice is a quadrilateral bounded by: (i) the physical light ray $\gamma_g$ from $S$ to $R$; (ii) two outgoing radial lines through $S$ and $R$; and (iii) an auxiliary arc (typically a circular orbit segment) closing the boundary. The Gauss-Bonnet theorem reads \cite{Gibbons:2008rj}
\begin{equation} \label{2.5}
\iint_D \mathcal{K}\,dS+\oint_{\partial D}k_g\,d\sigma+\sum_i \beta_i = 2\pi\chi(D),
\end{equation}
with Gaussian curvature $\mathcal K$, boundary geodesic curvature $k_g$, and jump (exterior) angles $\beta_i$ at the vertices.

In the setup adopted here, each boundary component is taken to be a geodesic in the optical metric, so $k_g=0$ along every segment. The only nontrivial bookkeeping is at the corners: at $R$ the jump angle is $\beta_R=\pi-\Psi_R$, while at $S$ it is $\beta_S=\Psi_S$ (orientation fixed by traversing $\partial D$ positively). With $\chi(D)=1$, Gauss-Bonnet gives
\begin{equation} \label{2.6}
\iint_D \mathcal{K}\,dS+\beta_S+\beta_R=\pi,
\end{equation}
which, combined with the definition of $\alpha$, yields the finite distance curvature formula
\begin{equation} \label{2.7}
\alpha=\iint_D \mathcal{K}\,dS+\phi_{RS}.
\end{equation}

\subsection{Li-type curvature primitive identity} \label{sec2.3}
A practical bottleneck in Gauss-Bonnet lensing is the area term
$\iint_D \mathcal K\,dS$, which in general requires an explicit parametrization of the domain $D$.
Li \emph{et al.} observed that for two-dimensional optical (or Jacobi) geometries with metric \cite{Li:2020wvn}
\begin{equation} \label{2.8}
dt^{2}=g_{rr}(r)\,dr^{2}+g_{\phi\phi}(r)\,d\phi^{2}
=\frac{B(r)}{A(r)}\,dr^{2}+\frac{C(r)}{A(r)}\,d\phi^{2},
\end{equation}
the integrand $\mathcal K\sqrt{\det g}$ is an exact radial derivative up to a constant.  Concretely, using
\begin{equation} \label{2.9}
\mathcal K=\frac{1}{\sqrt{\det g}}\left[
\frac{\partial}{\partial \phi}\!\left(\sqrt{\det g}\,g^{rr}\Gamma^{\phi}{}_{rr}\right)
-\frac{\partial}{\partial r}\!\left(\sqrt{\det g}\,g^{rr}\Gamma^{\phi}{}_{r\phi}\right)
\right],
\end{equation}
and the fact that all metric functions depend only on $r$, the $\partial_\phi$ term vanishes and one obtains
\begin{equation} \label{2.10}
\int \mathcal K\sqrt{\det g}\,dr
=-\sqrt{\det g}\,g^{rr}\Gamma^{\phi}{}_{r\phi}
=\frac{C A'-A C'}{2A\sqrt{BC}},
\end{equation}
where a prime denotes $d/dr$. We denote this antiderivative by
\begin{equation} \label{2.11}
\mathcal P(r)\;\equiv\;\int \mathcal K\sqrt{\det g}\,dr
=\frac{C A'-A C'}{2A\sqrt{BC}},
\end{equation}
keeping in mind that $\mathcal P$ is defined only up to an additive constant. 

Li’s operational simplification comes from choosing the Gauss-Bonnet domain to include a \emph{geodesic}
circular orbit $\gamma_{\rm co}$ as one boundary component.  For such an orbit, the geodesic curvature
vanishes and the normalization \cite{Li:2020wvn}
\begin{equation} \label{2.12}
\mathcal P(r_{\rm co})=0
\end{equation}
is natural (equivalently, one fixes the integration constant by subtracting $\mathcal P(r_{\rm co})$).
With this choice, the curvature-area contribution reduces to an integral \emph{along the physical ray}:
\begin{equation} \label{2.13}
\alpha
=\int_{\phi_S}^{\phi_R}\mathcal P\!\bigl(r(\phi)\bigr)\,d\phi+\phi_{RS}, 
\end{equation}
which is the starting point for our later radial-reformulation programme (the remaining dependence on
$r(\phi)$ is precisely what Section \ref{sec3} eliminates).

\section{Radial integral reformulation with complete derivation} \label{sec3}
\subsection{Null first integrals and orbit shape in differential form} \label{sec3.1}
We work on the equatorial plane $(\theta=\pi/2)$ of a static, spherically symmetric spacetime
\begin{equation}
ds^{2}=-A(r)\,dt^{2}+B(r)\,dr^{2}+C(r)\,d\phi^{2},
\qquad A(r)>0, \label{3.1}
\end{equation}
and parameterize the null geodesic by an affine parameter $\lambda$.
Stationarity and axisymmetry imply two first integrals,
\begin{equation}
E \equiv A(r)\,\frac{dt}{d\lambda},\qquad
L \equiv C(r)\,\frac{d\phi}{d\lambda}, \label{3.2}
\end{equation}
interpreted as the conserved energy and angular momentum (per unit mass) of the photon.
It is convenient to introduce the impact parameter
\begin{equation}
b\equiv \frac{L}{E}, \label{3.3}
\end{equation}
which is invariant under affine re-scalings of $\lambda$. Imposing the null condition $ds^{2}=0$ gives
\begin{equation}
0=-A\Bigl(\frac{dt}{d\lambda}\Bigr)^{2}+B\Bigl(\frac{dr}{d\lambda}\Bigr)^{2}
+C\Bigl(\frac{d\phi}{d\lambda}\Bigr)^{2}
=-\frac{E^{2}}{A}+B\Bigl(\frac{dr}{d\lambda}\Bigr)^{2}+\frac{L^{2}}{C}, \label{3.4}
\end{equation}
hence the radial first integral
\begin{equation}
\Bigl(\frac{dr}{d\lambda}\Bigr)^{2}
=\frac{E^{2}}{A(r)B(r)}-\frac{L^{2}}{B(r)C(r)}
=\frac{E^{2}}{A(r)B(r)}\left(1-\frac{A(r)b^{2}}{C(r)}\right). \label{3.5}
\end{equation}
Dividing the $\phi$-equation by the radial equation yields the orbit shape in differential form,
\begin{equation}\label{eq:dphidr_general}
\frac{d\phi}{dr}
=\frac{\frac{d\phi}{d\lambda}}{\frac{dr}{d\lambda}}
=\pm \frac{L/C(r)}{\sqrt{\frac{E^{2}}{A(r)B(r)}-\frac{L^{2}}{B(r)C(r)}}}
=\pm\frac{\sqrt{B(r)}}{\sqrt{C(r)}}
\left[\frac{C(r)}{A(r)b^{2}}-1\right]^{-1/2},
\end{equation}
which is the conversion factor used later to replace $\phi$-integration by $r$-integration.

A turning point (closest approach) occurs when $dr/d\lambda=0$, i.e. \cite{Claudel:2000yi}
\begin{equation}\label{eq:r0_condition}
\frac{C(r_0)}{A(r_0)b^{2}}-1=0
\quad\Longleftrightarrow\quad
C(r_0)=A(r_0)b^{2}.
\end{equation}
In the weak-deflection regime considered throughout this paper, we assume a \emph{single} turning point with
$r_S\ge r_0$ and $r_R\ge r_0$. Then $r(\lambda)$ is monotone decreasing from $r_S$ to $r_0$ on the incoming branch,
and monotone increasing from $r_0$ to $r_R$ on the outgoing branch, so that the sign in
Eq.~\eqref{eq:dphidr_general} flips across $r_0$ while $|d\phi/dr|$ remains continuous. 

\subsection{Starting point: Li one-dimensional reduction along the ray} \label{sec3.2}
Our radial reformulation begins from Li’s observation that, for the finite distance Gauss-Bonnet construction in an optical (or Jacobi) geometry, the curvature-area contribution can be reduced to a one-dimensional integral evaluated directly along the physical ray  \cite{Li:2020wvn}. Concretely, we consider a compact, simply connected domain $D$ on the optical manifold whose boundary consists of the null geodesic segment $\gamma_g$ connecting the source $S$ to the receiver $R$, two outgoing radial geodesics through $S$ and $R$, and a geodesic circular-orbit segment used to close the contour. For this choice, the geodesic curvature term vanishes on every boundary component, and the only contributions beyond the curvature integral arise from the corner (jump) angles, which are already absorbed into the finite distance definition of $\alpha$.

With these conditions, Li’s reduction yields
\begin{equation}
\alpha
=\int_{\phi_S}^{\phi_R}\mathcal P\!\bigl(r(\phi)\bigr)\,d\phi+\phi_{RS},
\label{3.8}
\end{equation}
where $\phi_{RS}\equiv\phi_R-\phi_S$, and $\mathcal P(r)$ is the curvature primitive obtained by integrating $\mathcal K\sqrt{\det g}$ with respect to $r$ in the optical metric. The primitive is defined only up to an additive constant; operationally, one fixes this constant by imposing a normalization on the auxiliary geodesic circular orbit (most commonly $\mathcal P(r_{\rm co})=0$), ensuring that the would-be lower limit contribution from the curvature-area integral cancels identically. The essential limitation of Eq. \eqref{3.8} is that it still depends on the orbit in the implicit form $r(\phi)$, so that direct evaluation typically requires either an explicit orbit inversion or repeated weak-field substitutions. Our goal in the next subsection is to eliminate this residual dependence by a controlled change of variables using $d\phi/dr$ and a careful split at the turning point.

\subsection{Variable change and splitting at the turning point} \label{sec3.3}
We start from the Li-reduced representation $\alpha$ along the ray as given by Eq. \eqref{3.8}, which holds when the Gauss-Bonnet domain is bounded by optical-geodesic segments so that boundary geodesic-curvature contributions vanish and the optical manifold is regular on the chosen domain.

Assume the light ray has a single turning point $r=r_0$ (closest approach) satisfying $C(r_0)=A(r_0)b^2$, with endpoints in the scattering configuration $r_S\ge r_0$ and $r_R\ge r_0$.
On each branch, $r$ is a good parameter and we may change variables in the $\phi$-integral via
$d\phi=(d\phi/dr)\,dr$. For null motion in the equatorial plane, the orbit-shape factor can be written as
\begin{equation}
\left|\frac{d\phi}{dr}\right|
=\frac{\sqrt{B(r)}}{\sqrt{C(r)}}
\left[\frac{C(r)}{A(r)b^{2}}-1\right]^{-1/2},
\qquad r\ge r_0,
\label{3.9}
\end{equation}
where the absolute value is taken so that the right-hand side is nonnegative on both branches.
The sign of $d\phi/dr$ flips at $r_0$ because $dr/d\lambda$ changes sign while $d\phi/d\lambda$ keeps a fixed sign along the trajectory. Consequently, the Li integral decomposes as
\begin{equation}
\int_{\phi_S}^{\phi_R}\mathcal P\!\bigl(r(\phi)\bigr)\,d\phi
=\int_{r_S}^{r_0}\mathcal P(r)\,\frac{d\phi}{dr}\,dr
+\int_{r_0}^{r_R}\mathcal P(r)\,\frac{d\phi}{dr}\,dr .
\label{3.10}
\end{equation}
Using $d\phi/dr=-|d\phi/dr|$ on the incoming branch ($r$ decreasing from $r_S$ to $r_0$) and
$d\phi/dr=+|d\phi/dr|$ on the outgoing branch ($r$ increasing from $r_0$ to $r_R$), Eq. \eqref{3.10} becomes
\begin{equation}
\int_{\phi_S}^{\phi_R}\mathcal P\!\bigl(r(\phi)\bigr)\,d\phi
=\int_{r_0}^{r_S}\mathcal P(r)\left|\frac{d\phi}{dr}\right|dr
+\int_{r_0}^{r_R}\mathcal P(r)\left|\frac{d\phi}{dr}\right|dr.
\label{3.11}
\end{equation}
Substituting Eq. \eqref{3.11} into Eq. \eqref{3.8} yields the desired turning-point radial form,
\begin{equation}
\alpha=
\int_{r_0}^{r_S}\mathcal P(r)\left|\frac{d\phi}{dr}\right|dr
+\int_{r_0}^{r_R}\mathcal P(r)\left|\frac{d\phi}{dr}\right|dr
+\phi_{RS}.
\label{3.12}
\end{equation}
The validity of Eq. \eqref{3.12} requires: (i) a single turning point so that the ray consists of exactly two monotone radial segments; (ii) absence of caustics in the weak-field domain so that the optical manifold description remains single-valued along the chosen ray; (iii) endpoints in the admissible region $A(r)>0$ with $r_S,r_R\ge r_0$; and (iv) existence of the auxiliary optical-geodesic boundaries used in the Gauss-Bonnet domain so that the Li reduction is applicable.

If there is no turning point (a monotonic ray), the decomposition is unnecessary: one may write
$\int_{\phi_S}^{\phi_R}\mathcal P\,d\phi=\int_{r_S}^{r_R}\mathcal P(r)\,(d\phi/dr)\,dr$ with the sign of
$d\phi/dr$ fixed throughout. The weak-deflection lensing configuration of interest typically exhibits a closest-approach radius $r_0$, and the split form \eqref{3.12} is precisely what makes subsequent weak-field evaluation modular, since each endpoint contributes through an integral over $[r_0,r_X]$ with $X\in\{S,R\}$.

\subsection{Normalization of the primitive and cancellation structure} \label{sec3.4}
The curvature primitive $\mathcal P(r)$ entering Eqs. \eqref{3.8} and \eqref{3.12} is defined only up to an additive constant, since it arises from an indefinite radial integration of $\mathcal K\sqrt{\det g}$. The invariant quantity is therefore the \emph{subtracted} primitive
\begin{equation}
\mathcal P_{\rm N}(r)\equiv \mathcal P(r)-\mathcal P(r_{\rm ref}),
\label{3.13}
\end{equation}
because $\mathcal P_{\rm N}$ is unchanged under $\mathcal P\mapsto \mathcal P+{\rm const}$. In Li’s construction one fixes $r_{\rm ref}$ by choosing a geodesic circular orbit $\gamma_{\rm co}$ and imposing $\mathcal P(r_{\rm co})=0$, so that the curvature-area term is expressed solely through evaluation along the physical ray.
With this understood, Eq. \eqref{3.12} is to be read with $\mathcal P\to\mathcal P_{\rm N}$, i.e.
\begin{equation}
\alpha=
\int_{r_0}^{r_S}\mathcal P_{\rm N}(r)\left|\frac{d\phi}{dr}\right|dr
+\int_{r_0}^{r_R}\mathcal P_{\rm N}(r)\left|\frac{d\phi}{dr}\right|dr
+\phi_{RS}.
\label{3.14}
\end{equation}

The cancellation structure becomes transparent in a weak-field expansion around a reference optical geometry. For the flat reference ($A=B=1$, $C=r^{2}$), the primitive constructed from Eq. \eqref{2.11} is a constant,
\begin{equation}
\mathcal P_{\rm flat}(r)=-1,
\label{3.15}
\end{equation}
so that $\mathcal P_{{\rm flat},{\rm N}}(r)=0$ for any finite choice of $r_{\rm ref}$, and the radial integrals in Eq. \eqref{3.14} vanish identically. The remaining $\phi_{RS}$ term is a purely kinematic endpoint contribution, and in the consistently defined finite distance deflection angle it cancels against the corresponding unlensed geometry. Consequently, Eq. \eqref{3.14} isolates the genuinely curvature-induced correction at the relevant perturbative order by removing the constant (background) piece at the integrand level.

\subsection{Specialization to Schwarzschild gauge} \label{sec3.5}
To make the radial reformulation operational, we now adopt the Schwarzschild gauge on the equatorial plane, defined by choosing the areal radius so that
\begin{equation}
ds^{2}=-A(r)\,dt^{2}+B(r)\,dr^{2}+r^{2}d\phi^{2}.
\label{3.16}
\end{equation}
In this gauge the orbit-shape factor in Eq. \eqref{3.9} becomes
\begin{equation}
\left|\frac{d\phi}{dr}\right|
=\frac{\sqrt{B(r)}}{r}\left[\frac{r^{2}}{A(r)b^{2}}-1\right]^{-1/2},
\qquad r\ge r_0,
\label{3.17}
\end{equation}
and the turning point is fixed by $r_0^{2}=A(r_0)b^{2}$. The Li primitive also simplifies because $C(r)=r^{2}$, so the combination $C A'-A C'$ reduces to $r^{2}A'(r)-2rA(r)$. Hence all curvature information entering Eq. \eqref{3.14} is encoded in $A(r)$ and $B(r)$ alone, with $r$ having a direct geometric meaning as the circumferential radius.

\section{Weak-field evaluation toolkit in the radial representation} \label{sec4}
In this section we develop a controlled perturbative scheme for evaluating the radial representation of the finite distance deflection angle derived in Section \ref{sec3}. The aim is to render the integrals explicit in a weak-field expansion while keeping the source and receiver at finite radii.

\subsection{Expansion scheme and working assumptions} \label{sec4.1}
We specialize to the Schwarzschild gauge,
\begin{equation}
ds^{2}=-A(r)\,dt^{2}+B(r)\,dr^{2}+r^{2}d\phi^{2},
\label{4.1}
\end{equation}
and assume a weak-field, static region in which $A(r)>0$ and the lensing trajectory has a single turning point $r=r_0$. We work in geometric units $G=c=1$ so that a mass parameter $M$ has dimensions of length and the dimensionless expansion variable is $\varepsilon(r)\equiv M/r$.

Our basic perturbative hypothesis is that $\varepsilon(r)\ll 1$ throughout the integration ranges relevant to finite distance lensing, namely $r\in[r_0,r_S]$ and $r\in[r_0,r_R]$. Accordingly, we parameterize the metric functions by asymptotic series of the form
\begin{equation}
A(r)=1+\sum_{n=1}^{N}a_n\,\varepsilon(r)^{n},\qquad
B(r)=1+\sum_{n=1}^{N}b_n\,\varepsilon(r)^{n},
\label{4.2}
\end{equation}
where the coefficients $a_n$ and $b_n$ are dimensionless constants determined by the underlying theory and gauge choice, and $N$ is the working truncation order. The normalization $A\to 1$ and $B\to 1$ as $r\to\infty$ ensures that the optical geometry approaches the flat reference needed for a clean subtraction of the constant piece of the curvature primitive.

The turning point is determined implicitly by the null first integrals through $r_0^{2}=A(r_0)b^{2}$, so $r_0$ itself is treated as a derived quantity with the same weak-field ordering. In practice, we expand every ingredient entering the radial formula for $\alpha$—the normalized primitive $\mathcal P_{\rm N}(r)$ and the orbit factor $|d\phi/dr|$—as a series in $\varepsilon(r)$, and then integrate term-by-term. The resulting expressions must be dimensionally consistent (each contribution to $\alpha$ is dimensionless) and are valid only within the domain where higher-order remainders are parametrically small, i.e. when $\max\{\varepsilon(r_0),\varepsilon(r_S),\varepsilon(r_R)\}\ll 1$ and the ray remains in the static region without encountering additional turning points.

\subsection{Reduction to standard  \texorpdfstring{$(r)$}{}-integral families} \label{sec4.2}
We now reduce the radial expression for the Li term in the deflection angle to a set of elementary integral families that can be evaluated once and reused at any weak-field order. In the Schwarzschild gauge, the absolute orbit factor follows from Eq. \eqref{3.17} together with the turning-point relation $b^{2}=r_{0}^{2}/A(r_{0})$, giving
\begin{equation}
\left|\frac{d\phi}{dr}\right|
=\frac{r_{0}\sqrt{A(r)B(r)}}{r\,\sqrt{\Delta(r)}},
\qquad
\Delta(r)\equiv r^{2}A(r_{0})-r_{0}^{2}A(r),
\qquad r\ge r_{0}.
\label{4.3}
\end{equation}
Accordingly, each endpoint contribution in Eq. \eqref{3.14} takes the form
\begin{equation}
\int_{r_0}^{r_X}\mathcal P_{\rm N}(r)\left|\frac{d\phi}{dr}\right|dr
=r_0\int_{r_0}^{r_X}\frac{\mathcal P_{\rm N}(r)\sqrt{A(r)B(r)}}{r\,\sqrt{\Delta(r)}}\,dr,
\qquad X\in\{S,R\}.
\label{4.4}
\end{equation}

To implement the weak-field scheme of Eq. \eqref{4.2} we write $A(r)=1+\delta A(r)$ and $B(r)=1+\delta B(r)$, where $\delta A,\delta B=\mathcal O(\varepsilon)$. Then $\Delta(r)$ decomposes as
\begin{equation}
\Delta(r)=(r^{2}-r_{0}^{2})+\delta\Delta(r),
\qquad
\delta\Delta(r)\equiv r^{2}\delta A(r_{0})-r_{0}^{2}\delta A(r),
\label{4.5}
\end{equation}
so that the square-root factor admits the binomial expansion
\begin{equation}
\frac{1}{\sqrt{\Delta(r)}}=
\frac{1}{\sqrt{r^{2}-r_{0}^{2}}}\left[
1-\frac{\delta\Delta(r)}{2(r^{2}-r_{0}^{2})}
+\frac{3}{8}\left(\frac{\delta\Delta(r)}{r^{2}-r_{0}^{2}}\right)^{2}
+\mathcal O\!\left(\left(\frac{\delta\Delta(r)}{r^{2}-r_{0}^{2}}\right)^{3}\right)
\right],
\label{4.6}
\end{equation}
truncated consistently at the same order $N$ as Eq. \eqref{4.2}. Similarly, we expand
$\sqrt{A(r)B(r)}=1+\mathcal O(\varepsilon)$ and the normalized primitive as
$\mathcal P_{\rm N}(r)=\sum_{n\ge 1}p_n\,\varepsilon(r)^{n}$, where the absence of a constant term reflects the subtraction described in Subsection~3.4.

After inserting these series into Eq. \eqref{4.4}, every term in the integrand becomes a rational function of $r$ multiplied by either $(r^{2}-r_{0}^{2})^{-1/2}$ or $(r^{2}-r_{0}^{2})^{-3/2}$ (and, at higher orders, higher odd half-powers). It is therefore sufficient to reduce the calculation to the families
\begin{equation}
\mathcal J_{m}^{(X)}\equiv \int_{r_{0}}^{r_{X}}\frac{dr}{r^{m}\sqrt{r^{2}-r_{0}^{2}}},
\qquad
\mathcal L_{m}^{(X)}\equiv \int_{r_{0}}^{r_{X}}\frac{dr}{r^{m}(r^{2}-r_{0}^{2})^{3/2}},
\qquad X\in\{S,R\},
\label{4.7}
\end{equation}
together with the trivial algebraic prefactors built from $r_{0}$ and the expansion coefficients.
All higher structures generated by Eq. \eqref{4.6} can be reduced to linear combinations of
$\mathcal J_{m}^{(X)}$ and $\mathcal L_{m}^{(X)}$ by rewriting $\delta\Delta(r)$ as a finite sum of monomials in
$r^{-1}$ and using partial-fraction manipulations in $r$.

In this way, the weak-field evaluation of the Li term becomes an exercise in (i) expanding the integrand in $\varepsilon$, (ii) collecting contributions by powers of $1/r$, and (iii) mapping each contribution to the pre-tabulated integral families in Eq. \eqref{4.7}, separately for $X=S$ and $X=R$.

\subsection{Endpoint geometry and finite distance angles} \label{sec4.3}
The finite distance definition of the deflection angle requires the local incidence angles $\Psi_S$ and $\Psi_R$ between the light-ray tangent and the outward radial direction at the endpoints, measured intrinsically in the optical geometry. In the Schwarzschild gauge, the optical line element on the equatorial plane is
\begin{equation}
dt^{2}=dl^{2}=\frac{B(r)}{A(r)}\,dr^{2}+\frac{r^{2}}{A(r)}\,d\phi^{2}.
\label{4.8}
\end{equation}
Let the ray be parameterized by the coordinate time $t$ so that the optical speed is unity, $g^{(\mathrm{opt})}_{ij}(dx^{i}/dt)(dx^{j}/dt)=1$. Using the conserved quantities of null motion, the angular velocity is
\begin{equation}
\frac{d\phi}{dt}=\frac{A(r)b}{r^{2}},
\label{4.9}
\end{equation}
where $b=L/E$ is the impact parameter. The radial velocity follows from the null condition,
\begin{equation}
\frac{dr}{dt}=\pm\sqrt{\frac{A(r)}{B(r)}}\sqrt{1-\frac{A(r)b^{2}}{r^{2}}}.
\label{4.10}
\end{equation}
Introduce an orthonormal optical frame $\hat e_{r}=\sqrt{A/B}\,\partial_{r}$ and $\hat e_{\phi}=\sqrt{A}\,r^{-1}\partial_{\phi}$. The unit tangent has components
\begin{equation}
v_{\hat r}\equiv \sqrt{\frac{B}{A}}\frac{dr}{dt}=\pm\sqrt{1-\frac{A(r)b^{2}}{r^{2}}},\qquad
v_{\hat\phi}\equiv \frac{r}{\sqrt{A}}\frac{d\phi}{dt}=\frac{b\sqrt{A(r)}}{r},
\label{4.11}
\end{equation}
which satisfy $v_{\hat r}^{2}+v_{\hat\phi}^{2}=1$ by construction. We therefore define $\Psi\in[0,\pi]$ by
\begin{equation}
\sin\Psi=\frac{b\sqrt{A(r)}}{r},\qquad
\cos\Psi=\sqrt{1-\frac{A(r)b^{2}}{r^{2}}},
\label{4.12}
\end{equation}
with the sign of $v_{\hat r}$ distinguishing incoming from outgoing motion but not affecting $\Psi$. Evaluating Eq. \eqref{4.12} at $r=r_S$ and $r=r_R$ yields $\Psi_S$ and $\Psi_R$ in terms of $(b,A)$ and the endpoint radii, making the finite distance angle contribution explicit once $b$ (or equivalently $r_0$) is fixed.

\section{Worked examples} \label{sec5}
We now illustrate the radial representation by evaluating the finite distance deflection angle in standard metrics and verifying that the known limiting behaviors are recovered.

\subsection{Schwarzschild spacetime at finite distance} \label{sec5.1}
For the Schwarzschild spacetime, we take
\begin{equation}
A(r)=1-\frac{2M}{r},\qquad B(r)=\frac{1}{1-\frac{2M}{r}},
\label{5.1}
\end{equation}
in the static region $r>2M$. Substituting Eq. \eqref{5.1} into the Schwarzschild-gauge primitive yields the closed form
\begin{equation}
\mathcal P(r)=\frac{-1+\frac{3M}{r}}{\sqrt{1-\frac{2M}{r}}}.
\label{5.2}
\end{equation}
The flat reference gives $\mathcal P_{\rm flat}=-1$, so the normalized primitive is
\begin{equation}
\mathcal P_{\rm N}(r)=\mathcal P(r)+1=\frac{2M}{r}+\mathcal O\!\left(\frac{M^{2}}{r^{2}}\right).
\label{5.3}
\end{equation}
At the same accuracy, the orbit factor in Eq. \eqref{4.3} reduces to its leading geometric piece,
\begin{equation}
\left|\frac{d\phi}{dr}\right|
=\frac{r_0}{r\sqrt{r^{2}-r_0^{2}}}
+\mathcal O\!\left(\frac{M}{r}\right),
\qquad r\ge r_0,
\label{5.4}
\end{equation}
where $r_0$ is the turning point. Inserting Eqs. \eqref{5.3} and \eqref{5.4} into the radial Li term in Eq. \eqref{3.14} and integrating term-by-term gives, for each endpoint $X\in\{S,R\}$,
\begin{equation}
\int_{r_0}^{r_X}\mathcal P_{\rm N}(r)\left|\frac{d\phi}{dr}\right|dr
=\frac{2M}{r_0}\sqrt{1-\frac{r_0^{2}}{r_X^{2}}}
+\mathcal O\!\left(\frac{M^{2}}{r_0^{2}}\right).
\label{5.5}
\end{equation}
Therefore, to leading weak-field order, the finite distance deflection angle is
\begin{equation}
\alpha
=\frac{2M}{r_0}\left(\sqrt{1-\frac{r_0^{2}}{r_R^{2}}}
+\sqrt{1-\frac{r_0^{2}}{r_S^{2}}}\right)
+\mathcal O\!\left(\frac{M^{2}}{r_0^{2}}\right),
\label{5.6}
\end{equation}
which matches the finite distance formulation based on endpoint angles in the optical geometry.
In the asymptotic limit $r_S,r_R\to\infty$, the square roots tend to unity and Eq. \eqref{5.6} reduces to
$\alpha=4M/r_0+\mathcal O(M^{2}/r_0^{2})$. Identifying $r_0=b+\mathcal O(M)$ then reproduces the textbook leading term $4M/b$ obtained in both the standard approach and the Gauss-Bonnet optical-geometry method.

\subsection{Kottler spacetime Schwarzschild de Sitter} \label{sec5.2}
We take the Kottler (Schwarzschild-de Sitter) line element in Schwarzschild gauge,
\begin{equation}
A(r)=1-\frac{2M}{r}-\frac{\Lambda r^{2}}{3},\qquad
B(r)=\frac{1}{A(r)},\qquad
C(r)=r^{2},
\label{5.7}
\end{equation}
with $M>0$ and (for definiteness) $\Lambda>0$.
The optical geometry relevant to our construction is restricted to the static region where $A(r)>0$. For $\Lambda>0$ and sufficiently small $M$, $A(r)$ has two positive simple zeros corresponding to a black-hole horizon $r_b$ and a cosmological horizon $r_c$, and the static patch is $r\in(r_b,r_c)$. A necessary condition for this two-horizon structure is the familiar bound $M<1/(3\sqrt{\Lambda})$ (Nariai limit at equality) \cite{Ginsparg:1982rs,Israel:1966rt}.
Throughout this example we assume $r_0,r_S,r_R\in(r_b,r_c)$ and, for weak lensing, the additional smallness conditions
$M/r\ll 1$ and $\Lambda r^{2}\ll 1$ on the full integration ranges $r\in[r_0,r_S]$ and $r\in[r_0,r_R]$.

In Schwarzschild gauge with $B=1/A$ and $C=r^{2}$, the Li curvature primitive may be written as
\begin{equation}
\mathcal P(r)=\frac{rA'(r)-2A(r)}{2\sqrt{A(r)}}.
\label{5.8}
\end{equation}
Substituting Eq. \eqref{5.7} gives $rA'(r)-2A(r)=-2+6M/r$, so
\begin{equation}
\mathcal P(r)=\frac{-1+\frac{3M}{r}}{\sqrt{1-\frac{2M}{r}-\frac{\Lambda r^{2}}{3}}}.
\label{5.9}
\end{equation}
The geodesic circular-orbit normalization used in Li’s reduction is automatically compatible with Kottler: the circular null orbit is fixed by $rA'(r)-2A(r)=0$, hence $r_{\rm co}=3M$ independent of $\Lambda$, and Eq. \eqref{5.9} satisfies $\mathcal P(3M)=0$ without any additional constant shift.

To obtain a finite distance weak-deflection formula, we expand Eq. \eqref{5.9} to linear order in $M/r$ and $\Lambda r^{2}$,
\begin{equation}
\mathcal P(r)=-1+\frac{2M}{r}-\frac{\Lambda r^{2}}{6}
+\mathcal O\!\left(\frac{M^{2}}{r^{2}},\,\Lambda^{2}r^{4},\,\Lambda Mr\right),
\label{5.10}
\end{equation}
and use the leading orbit factor,
\begin{equation}
\left|\frac{d\phi}{dr}\right|=\frac{r_0}{r\sqrt{r^{2}-r_0^{2}}}
+\mathcal O\!\left(\frac{M}{r}\right)+\mathcal O\!\left(\Lambda r^{2}\right),
\qquad r\ge r_0,
\label{5.11}
\end{equation}
since multiplying $\mathcal P(r)$ by the $\mathcal O(M)$ or $\mathcal O(\Lambda)$ corrections in $|d\phi/dr|$ would only contribute beyond the order retained. The constant piece $-1$ in Eq. \eqref{5.10} integrates to
$-\phi_{RS}^{(0)}$, where the Euclidean (zeroth-order) coordinate separation along the straight-line proxy is
$\phi_{RS}^{(0)}=\arccos(r_0/r_S)+\arccos(r_0/r_R)$, so it cancels against the explicit $+\phi_{RS}$ term in the Li formula when $\phi_{RS}$ is evaluated at the same order.
The surviving contributions are therefore sourced by the $2M/r$ and $-\Lambda r^{2}/6$ terms in Eq. \eqref{5.10}. Carrying out the remaining radial integrals yields, for each endpoint $X\in\{S,R\}$,
\begin{equation}
\int_{r_0}^{r_X}\left(\frac{2M}{r}\right)\frac{r_0\,dr}{r\sqrt{r^{2}-r_0^{2}}}
=\frac{2M}{r_0}\sqrt{1-\frac{r_0^{2}}{r_X^{2}}},
\qquad
\int_{r_0}^{r_X}\left(-\frac{\Lambda r^{2}}{6}\right)\frac{r_0\,dr}{r\sqrt{r^{2}-r_0^{2}}}
=-\frac{\Lambda r_0}{6}\sqrt{r_X^{2}-r_0^{2}}.
\label{5.12}
\end{equation}
Summing the source and receiver pieces gives the finite distance weak deflection angle,
\begin{equation}
\alpha
=\frac{2M}{r_0}\left(\sqrt{1-\frac{r_0^{2}}{r_S^{2}}}+\sqrt{1-\frac{r_0^{2}}{r_R^{2}}}\right)
-\frac{\Lambda r_0}{6}\left(\sqrt{r_S^{2}-r_0^{2}}+\sqrt{r_R^{2}-r_0^{2}}\right)
+\mathcal O\!\left(\frac{M^{2}}{r_0^{2}},\,\Lambda^{2}r_0^{4},\,\Lambda Mr_0\right).
\label{5.13}
\end{equation}
If we replace $r_0$ by the impact parameter $b$ at this order and define $\chi_X\equiv\sqrt{1-b^{2}/r_X^{2}}$, Eq. \eqref{5.13} becomes
$\alpha=\frac{2M}{b}(\chi_S+\chi_R)-\frac{\Lambda b}{6}(r_S\chi_S+r_R\chi_R)
+\mathcal O\!\left(\frac{M^{2}}{b^{2}},\,\Lambda^{2}b^{4},\,\Lambda Mb\right)$, which makes the finite distance scaling of the $\Lambda$-correction explicit. The limiting checks follow immediately: letting $\Lambda\to 0$ reduces Eq. \eqref{5.13} to the Schwarzschild finite distance result, while weak-field consistency requires $r_0,r_S,r_R\ll r_c\sim \sqrt{3/\Lambda}$ so that the ray remains inside the static region and the perturbative ordering is maintained.

\subsection{Black hole surrounded by perfect fluid dark matter with finite halo endpoints} \label{sec5.3}
We consider the static, spherically symmetric \emph{perfect fluid dark matter} (PFDM) deformation of the Schwarzschild gauge, in which the lapse function acquires a logarithmic correction. In the (halo) interior we take
\begin{equation}
A_{\rm in}(r)=1-\frac{2M}{r}+\frac{\epsilon}{r}\ln\!\Bigl(\frac{r}{|\epsilon|}\Bigr),\qquad
B_{\rm in}(r)=\frac{1}{A_{\rm in}(r)},\qquad C(r)=r^{2},
\label{5.14}
\end{equation}
where $M$ is the black-hole mass parameter and $\epsilon$ is the PFDM intensity parameter (both with dimensions of length in $G=c=1$ units). This functional form is standard in the PFDM black-hole literature \cite{Rahaman:2010xs,Xu:2017bpz,Hou:2018avu,Hou:2018bar,Haroon:2018ryd,Gao:2023ltr,Ahmed:2025ojg}.
To model a \emph{finite} halo we introduce a halo radius $r_h$ and adopt a minimal exterior completion that preserves continuity of $A(r)$ at $r=r_h$ by freezing the logarithm at its halo-edge value,
\begin{equation}
A_{\rm out}(r)=1-\frac{2M}{r}+\frac{\epsilon}{r}\ln\!\Bigl(\frac{r_h}{|\epsilon|}\Bigr),\qquad
B_{\rm out}(r)=\frac{1}{A_{\rm out}(r)},\qquad C(r)=r^{2}.
\label{5.15}
\end{equation}
We classify endpoints by whether $r_S$ and $r_R$ lie inside ($r<r_h$) or outside ($r>r_h$). The mixed configurations are $(S\in,R\notin)$ with $r_S<r_h<r_R$ and $(S\notin,R\in)$ with $r_R<r_h<r_S$, and we assume a weak-deflection trajectory whose turning point lies inside the halo, $b<r_h$, where $b$ is the impact parameter.

Before continuing, we state some remarks on the halo-edge matching at $r=r_h$. The finite-halo prescription in Eqs.\eqref{5.14}-\eqref{5.15} makes $A(r)$ continuous at $r=r_h$ but, as written, it need not be $C^{1}$ there. 
Since the optical curvature $\mathcal K$ involves derivatives of the optical metric, a lack of differentiability at $r_h$ may be interpreted as an idealized thin transition layer (and can be associated with a distributional curvature contribution at the matching surface in the strict sharp-limit picture). To avoid this ambiguity while retaining the same physical content, one may regard Eqs. \eqref{5.14}-\eqref{5.15} as the $\varepsilon\to0$ limit of a smooth finite-halo profile.
For example, introduce a smooth window function $W_\varepsilon(r)$ with width $\varepsilon\ll r_h$,
\begin{equation}
W_\varepsilon(r)=\frac12\left[1-\tanh\!\left(\frac{r-r_h}{\varepsilon}\right)\right],
\qquad
W_\varepsilon(r)\to
\begin{cases}
1,& r<r_h,\\
0,& r>r_h,
\end{cases}
\ \text{as } \varepsilon\to0,
\label{eq:Weps}
\end{equation}
and replace the sharp exterior freezing by the smooth interpolation
\begin{equation}
\ln r\ \longrightarrow\ W_\varepsilon(r)\,\ln r+\bigl(1-W_\varepsilon(r)\bigr)\,\ln r_h .
\label{eq:log_smooth}
\end{equation}
Then $A_\varepsilon(r)$ is smooth for every $\varepsilon>0$, and Eqs. \eqref{5.14}-\eqref{5.15} are recovered as $\varepsilon\to0$. In the Li reduction, the only effect of the transition layer in the sharp limit is a finite matching contribution equal to the jump of the Li primitive across $r_h$.
Equivalently, when the radial integrals are split at $r_h$, the result should be interpreted with one-sided evaluations at $r_h$ (from the interior and exterior) kept distinct; this is precisely the $\varepsilon\to0$ limit of a smooth finite-halo profile, and no additional shell term is missed by our procedure.

The radial-GB pipeline for mixed endpoints proceeds exactly as in Section \ref{sec3}, with the only change being that the Li integrals are split at $r=r_h$ whenever the ray traverses the matching surface. To leading weak-field order we may set the turning-point radius $r_0=b$ and use the zeroth-order orbit factor
\begin{equation}
\left|\frac{d\phi}{dr}\right|=\frac{b}{r\sqrt{r^{2}-b^{2}}}.
\label{5.16}
\end{equation}
The Li contribution for an endpoint at $r=r_X$ is then $\int_{b}^{r_X}\mathcal P_{\rm N}(r)\,|d\phi/dr|\,dr$, with $\mathcal P_{\rm N}$ evaluated from the appropriate branch of the metric; for example, in the $(S\in,R\notin)$ case the receiver-side integral becomes
\begin{equation}
\int_{b}^{r_h}\!\Bigl(\mathcal P_{\rm N}\Bigr)_{\rm in}\left|\frac{d\phi}{dr}\right|dr
\;+\;
\int_{r_h}^{r_R}\!\Bigl(\mathcal P_{\rm N}\Bigr)_{\rm out}\left|\frac{d\phi}{dr}\right|dr,
\label{5.17}
\end{equation}
while the source-side integral is purely interior. The explicit $+\phi_{RS}$ term in the Li formula cancels the integrated constant ($-1$) piece of $\mathcal P$ at the same perturbative order, so the weak deflection is governed by the normalized primitive $\mathcal P_{\rm N}=\mathcal P-\mathcal P_{\rm flat}$ with $\mathcal P_{\rm flat}=-1$.

For $B=1/A$ and $C=r^{2}$, the primitive is $\mathcal P=(rA'-2A)/(2\sqrt{A})$. Expanding to linear order in $\delta A\equiv A-1$ gives $\mathcal P_{\rm N}(r)=(r\delta A'(r)-\delta A(r))/2+\mathcal O(\delta A^{2})$, so that the universal mass term is $\mathcal P_{{\rm N},M}(r)=2M/r$. For the PFDM part we obtain
\begin{equation}
\Bigl(\mathcal P_{\rm N}\Bigr)_{\rm in}(r)=\frac{2M}{r}
+\frac{\epsilon}{2r}\!\left[1-2\ln\!\Bigl(\frac{r}{|\epsilon|}\Bigr)\right],
\qquad
\Bigl(\mathcal P_{\rm N}\Bigr)_{\rm out}(r)=\frac{2M}{r}
-\frac{\epsilon}{r}\ln\!\Bigl(\frac{r_h}{|\epsilon|}\Bigr),
\label{5.18}
\end{equation}
valid when $M/r\ll 1$ and $|\epsilon|/r\ll 1$ on the full integration ranges.

It is convenient to introduce the finite distance endpoint factor
\begin{equation}
\chi(x)\equiv \sqrt{1-\frac{b^{2}}{x^{2}}},\qquad x\ge b,
\label{5.19}
\end{equation}
and denote $\chi_S=\chi(r_S)$, $\chi_R=\chi(r_R)$, and $\chi_h=\chi(r_h)$. The leading deflection angle admits the decomposition
\begin{equation}
\alpha^{\rm mixed}=\alpha_{M}^{\rm mixed}+\alpha_{\epsilon}^{\rm mixed}
+\mathcal O\!\left(\frac{M^{2}}{b^{2}},\,\frac{\epsilon^{2}}{b^{2}},\,\frac{M|\epsilon|}{b^{2}}\right),
\label{5.20}
\end{equation}
where the mass contribution is universal (independent of the halo truncation at this order),
\begin{equation}
\alpha_{M}^{\rm mixed}=\frac{2M}{b}\left(\chi_S+\chi_R\right).
\label{5.21}
\end{equation}
The PFDM correction depends on the endpoint taxonomy. For $(S\in,R\notin)$ with $r_S<r_h<r_R$ and $b<r_h$, the split at $r_h$ yields
\begin{equation}
\alpha_{\epsilon}^{(S\in,R\notin)}=
\frac{\epsilon}{2b}\,\Xi_{\rm in}(r_S)
+\frac{\epsilon}{2b}\,\Xi_{\rm in}(r_h)
-\frac{\epsilon}{b}\ln\!\Bigl(\frac{r_h}{|\epsilon|}\Bigr)\left(\chi_R-\chi_h\right),
\label{5.22}
\end{equation}
where the interior closed form is
\begin{equation}
\Xi_{\rm in}(x)\equiv
-\chi(x)\!\left[1+2\ln\!\Bigl(\frac{b}{|\epsilon|}\Bigr)\right]
+2\ln\!\bigl(1+\chi(x)\bigr)+2\bigl(\chi(x)-1\bigr)\ln\!\Bigl(\frac{b}{x}\Bigr).
\label{5.23}
\end{equation}
For the opposite mixed configuration $(S\notin,R\in)$ with $r_R<r_h<r_S$ and $b<r_h$, the same building blocks give
\begin{equation}
\alpha_{\epsilon}^{(S\notin,R\in)}=
\frac{\epsilon}{2b}\,\Xi_{\rm in}(r_R)
+\frac{\epsilon}{2b}\,\Xi_{\rm in}(r_h)
-\frac{\epsilon}{b}\ln\!\Bigl(\frac{r_h}{|\epsilon|}\Bigr)\left(\chi_S-\chi_h\right).
\label{5.24}
\end{equation}
Equations \eqref{5.22}–\eqref{5.24} are the promised closed-form weak deflection formulas for mixed endpoints: they preserve finite distance information through the exact endpoint factors $\chi_S,\chi_R,\chi_h$ while isolating the halo contribution through the single matching scale $r_h$ and the logarithmic structure inherited from Eq. \eqref{5.14}. These expressions are dimensionally consistent (each term scales as $\epsilon/b$) and reduce smoothly to the Schwarzschild finite distance result when $\epsilon\to 0$, while their domain of validity is set by $M/b\ll 1$, $|\epsilon|/b\ll 1$, and the requirement that all radii involved remain in the static region of the interior and exterior metrics.

\section{Conclusion} \label{sec6}
This work reformulates finite distance gravitational lensing in optical geometry into a radial integral framework that is both algorithmic and flexible. Starting from the Gauss-Bonnet characterization of the finite distance deflection angle, the standard difficulty is the explicit evaluation of the curvature-area term over a nontrivial domain on the optical manifold. The Li-type curvature primitive identity provides a decisive simplification by converting the curvature-area contribution into a one-dimensional integral evaluated \emph{along the physical light ray}. However, that representation still carries an implicit dependence on the orbit in the form $r(\phi)$, which typically reintroduces iterative orbit inversion and repeated weak-field substitutions.

The central contribution of this paper is the complete derivation of a purely radial representation that eliminates the residual orbit dependence at the integrand level. The key steps are: (i) expressing the null geodesic by first integrals in a general static, spherically symmetric spacetime; (ii) converting the Li line integral $\int \mathcal P(r(\phi))\,d\phi$ into $\int \mathcal P(r)\,(d\phi/dr)\,dr$; and (iii) handling the sign change of $d\phi/dr$ by splitting the trajectory at the turning point $r=r_0$. The resulting formula decomposes the deflection into endpoint contributions over $[r_0,r_S]$ and $[r_0,r_R]$, plus the finite distance angular bookkeeping encoded in $\phi_{RS}$. This decomposition clarifies the geometric content: the lensing correction is controlled by local curvature data through $\mathcal P_{\rm N}(r)$, while the endpoint dependence is explicit and modular.

A nontrivial technical point is normalization of the primitive. Because $\mathcal P$ is defined up to an additive constant, naive implementations can inadvertently mix physical curvature contributions with background (reference) pieces. We showed how a normalized primitive $\mathcal P_{\rm N}$ isolates curvature-induced effects by removing the constant component (equivalently fixed by a reference radius or by the circular-orbit normalization adopted in Li’s construction). In weak-field practice, this normalization enforces an integrand-level cancellation that prevents spurious flat contributions from contaminating the finite distance deflection.

Section \ref{sec4} developed a practical toolkit for weak-field evaluation: after expanding $A(r)$ and $B(r)$, the radial representation reduces to a small set of standard integral families in $(r^2-r_0^2)^{-1/2}$ and its higher odd half-powers. This yields a reusable pipeline: once the expansions of $\mathcal P_{\rm N}(r)$ and $|d\phi/dr|$ are generated to a desired order, the deflection follows from tabulated integrals with explicit dependence on endpoint radii. Importantly, the endpoint geometry is handled intrinsically in the optical metric, and the finite distance incidence angles admit compact expressions in terms of $A(r)$, $b$, and the endpoint radii. Thus, the method remains well-adapted to observationally relevant configurations where neither the source nor the receiver is at infinity.

The worked examples serve two purposes. First, they validate the formalism by reproducing known limiting cases: Schwarzschild at finite distance reduces to the standard $4M/b$ result as $r_S,r_R\to\infty$, and the Kottler example illustrates how the pipeline treats additional scales and static-patch restrictions. Second, they demonstrate that the approach naturally accommodates piecewise metrics and \emph{mixed endpoint} configurations. The PFDM finite-halo model is representative: splitting the Li integral at the halo boundary produces closed-form weak-deflection expressions that track whether each endpoint lies inside or outside the halo. This is precisely the kind of bookkeeping that is cumbersome in area-integral formulations and becomes straightforward in the radial representation.

Several limitations and extensions are immediate. The present derivation assumes a single turning point and excludes caustics within the chosen optical domain; strong-deflection regimes with multiple windings require additional domain management and, potentially, a modified splitting strategy. While we specialized to the Schwarzschild gauge for operational simplicity, the underlying variable-change argument is not tied to that gauge; extending the explicit integral families to alternative radial gauges is straightforward but algebraically heavier. A particularly important direction is rotation: extending the method to stationary axisymmetric spacetimes requires handling a Randers-type optical geometry (or a Finslerian formulation), where the geodesic-curvature bookkeeping changes qualitatively. Another natural generalization is dispersive media (plasma lensing), where the optical metric becomes frequency-dependent and the primitive structure must be revisited.

In conclusion, the radial integral reformulation provides a clean interface between differential-geometric lensing (Gauss-Bonnet in optical geometry) and practical perturbative evaluation at finite distance. It converts curvature-based lensing into a transparent endpoint-wise computation, supports mixed-domain geometries such as finite halos, and yields closed-form weak-field expressions with clear validity conditions. This framework is intended to function as a reusable module: once $\mathcal P_{\rm N}$ and $|d\phi/dr|$ are specified for a model, finite distance lensing observables follow systematically.

\acknowledgments
A. \"O. and R. P. would like to acknowledge networking support of the COST Action CA21106 - COSMIC WISPers in the Dark Universe: Theory, astrophysics and experiments (CosmicWISPers), the COST Action CA22113 - Fundamental challenges in theoretical physics (THEORY-CHALLENGES), the COST Action CA21136 - Addressing observational tensions in cosmology with systematics and fundamental physics (CosmoVerse), the COST Action CA23130 - Bridging high and low energies in search of quantum gravity (BridgeQG), and the COST Action CA23115 - Relativistic Quantum Information (RQI) funded by COST (European Cooperation in Science and Technology). A. \"O. also thanks to EMU, TUBITAK, ULAKBIM (Turkiye) and SCOAP3 (Switzerland) for their support.

\bibliography{ref}

@article{Gibbons:2008rj,
    author = "Gibbons, G. W. and Werner, M. C.",
    title = "{Applications of the Gauss-Bonnet theorem to gravitational lensing}",
    eprint = "0807.0854",
    archivePrefix = "arXiv",
    primaryClass = "gr-qc",
    doi = "10.1088/0264-9381/25/23/235009",
    journal = "Class. Quant. Grav.",
    volume = "25",
    pages = "235009",
    year = "2008"
}

@article{Ishihara:2016sfv,
    author = "Ishihara, Asahi and Suzuki, Yusuke and Ono, Toshiaki and Asada, Hideki",
    title = "{Finite-distance corrections to the gravitational bending angle of light in the strong deflection limit}",
    eprint = "1612.04044",
    archivePrefix = "arXiv",
    primaryClass = "gr-qc",
    doi = "10.1103/PhysRevD.95.044017",
    journal = "Phys. Rev. D",
    volume = "95",
    number = "4",
    pages = "044017",
    year = "2017"
}

@article{Li:2024ujw,
    author = "Li, Zonghai",
    title = "{Gravitational lensing using Werner{\textquoteright}s method in Cartesian-like coordinates}",
    eprint = "2404.19658",
    archivePrefix = "arXiv",
    primaryClass = "gr-qc",
    doi = "10.1103/PhysRevD.111.084017",
    journal = "Phys. Rev. D",
    volume = "111",
    number = "8",
    pages = "084017",
    year = "2025"
}

@article{Allendoerfer_1943,
  author    = {Allendoerfer, Carl B. and Weil, André},
  journal   = {Transactions of the American Mathematical Society},
  title     = {The Gauss-Bonnet theorem for Riemannian polyhedra},
  year      = {1943},
  issn      = {1088-6850},
  number    = {1},
  pages     = {101--129},
  volume    = {53},
  doi       = {10.1090/S0002-9947-1943-0007627-9},
  publisher = {American Mathematical Society (AMS)},
}

@article{Li:2020wvn,
    author = {Li, Zonghai and Zhang, Guodong and {\"O}vg{\"u}n, Ali},
    title = "{Circular Orbit of a Particle and Weak Gravitational Lensing}",
    eprint = "2006.13047",
    archivePrefix = "arXiv",
    primaryClass = "gr-qc",
    doi = "10.1103/PhysRevD.101.124058",
    journal = "Phys. Rev. D",
    volume = "101",
    number = "12",
    pages = "124058",
    year = "2020"
}

@article{Li:2019qyb,
    author = "Li, Zonghai and Jia, Junji",
    title = "{The finite-distance gravitational deflection of massive particles in stationary spacetime: a Jacobi metric approach}",
    eprint = "1912.05194",
    archivePrefix = "arXiv",
    primaryClass = "gr-qc",
    doi = "10.1140/epjc/s10052-020-7665-8",
    journal = "Eur. Phys. J. C",
    volume = "80",
    number = "2",
    pages = "157",
    year = "2020"
}

@article{Rindler:2007zz,
    author = "Rindler, Wolfgang and Ishak, Mustapha",
    title = "{Contribution of the cosmological constant to the relativistic bending of light revisited}",
    eprint = "0709.2948",
    archivePrefix = "arXiv",
    primaryClass = "astro-ph",
    doi = "10.1103/PhysRevD.76.043006",
    journal = "Phys. Rev. D",
    volume = "76",
    pages = "043006",
    year = "2007"
}

@article{Gibbons:2008zi,
    author = "Gibbons, G. W. and Herdeiro, C. A. R. and Warnick, C. M. and Werner, M. C.",
    title = "{Stationary Metrics and Optical Zermelo-Randers-Finsler Geometry}",
    eprint = "0811.2877",
    archivePrefix = "arXiv",
    primaryClass = "gr-qc",
    doi = "10.1103/PhysRevD.79.044022",
    journal = "Phys. Rev. D",
    volume = "79",
    pages = "044022",
    year = "2009"
}

@article{Crisnejo:2018uyn,
    author = "Crisnejo, Gabriel and Gallo, Emanuel",
    title = "{Weak lensing in a plasma medium and gravitational deflection of massive particles using the Gauss-Bonnet theorem. A unified treatment}",
    eprint = "1804.05473",
    archivePrefix = "arXiv",
    primaryClass = "gr-qc",
    doi = "10.1103/PhysRevD.97.124016",
    journal = "Phys. Rev. D",
    volume = "97",
    number = "12",
    pages = "124016",
    year = "2018"
}

@article{Arakida:2017hrm,
    author = "Arakida, Hideyoshi",
    title = "{Light deflection and Gauss{\textendash}Bonnet theorem: definition of total deflection angle and its applications}",
    eprint = "1708.04011",
    archivePrefix = "arXiv",
    primaryClass = "gr-qc",
    doi = "10.1007/s10714-018-2368-2",
    journal = "Gen. Rel. Grav.",
    volume = "50",
    number = "5",
    pages = "48",
    year = "2018"
}

@article{Gibbons:2015qja,
    author = "Gibbons, G. W.",
    title = "{The Jacobi-metric for timelike geodesics in static spacetimes}",
    eprint = "1508.06755",
    archivePrefix = "arXiv",
    primaryClass = "gr-qc",
    doi = "10.1088/0264-9381/33/2/025004",
    journal = "Class. Quant. Grav.",
    volume = "33",
    number = "2",
    pages = "025004",
    year = "2016"
}

@article{Ono:2019hkw,
    author = "Ono, Toshiaki and Asada, Hideki",
    title = "{The effects of finite distance on the gravitational deflection angle of light}",
    eprint = "1906.02414",
    archivePrefix = "arXiv",
    primaryClass = "gr-qc",
    doi = "10.3390/universe5110218",
    journal = "Universe",
    volume = "5",
    number = "11",
    pages = "218",
    year = "2019"
}

@article{Chern_1944,
  author    = {Chern, Shiing-Shen},
  journal   = {The Annals of Mathematics},
  title     = {A Simple Intrinsic Proof of the Gauss-Bonnet Formula for Closed Riemannian Manifolds},
  year      = {1944},
  issn      = {0003-486X},
  month     = oct,
  number    = {4},
  pages     = {747},
  volume    = {45},
  doi       = {10.2307/1969302},
  publisher = {JSTOR},
}

@article{Jia:2020xbc,
    author = "Jia, Junji",
    title = "{The perturbative approach for the weak deflection angle}",
    eprint = "2001.02038",
    archivePrefix = "arXiv",
    primaryClass = "gr-qc",
    doi = "10.1140/epjc/s10052-020-7796-y",
    journal = "Eur. Phys. J. C",
    volume = "80",
    number = "3",
    pages = "242",
    year = "2020"
}

@book{perlick2003ray,
  title={Ray Optics, Fermat’s Principle, and Applications to General Relativity},
  author={Perlick, V.},
  isbn={9783540466628},
  series={Lecture Notes in Physics Monographs},
  year={2003},
  publisher={Springer Berlin Heidelberg}
}

@article{Adler:2022qtb,
    author = "Adler, Stephen L. and Virbhadra, K. S.",
    title = "{Cosmological constant corrections to the photon sphere and black hole shadow radii}",
    eprint = "2205.04628",
    archivePrefix = "arXiv",
    primaryClass = "gr-qc",
    doi = "10.1007/s10714-022-02976-7",
    journal = "Gen. Rel. Grav.",
    volume = "54",
    number = "8",
    pages = "93",
    year = "2022"
}

@article{Virbhadra:2008ws,
    author = "Virbhadra, K. S.",
    title = "{Relativistic images of Schwarzschild black hole lensing}",
    eprint = "0810.2109",
    archivePrefix = "arXiv",
    primaryClass = "gr-qc",
    doi = "10.1103/PhysRevD.79.083004",
    journal = "Phys. Rev. D",
    volume = "79",
    pages = "083004",
    year = "2009"
}

@article{Claudel:2000yi,
    author = "Claudel, Clarissa-Marie and Virbhadra, K. S. and Ellis, G. F. R.",
    title = "{The Geometry of photon surfaces}",
    eprint = "gr-qc/0005050",
    archivePrefix = "arXiv",
    doi = "10.1063/1.1308507",
    journal = "J. Math. Phys.",
    volume = "42",
    pages = "818--838",
    year = "2001"
}

@article{Virbhadra:1999nm,
    author = "Virbhadra, K. S. and Ellis, George F. R.",
    title = "{Schwarzschild black hole lensing}",
    eprint = "astro-ph/9904193",
    archivePrefix = "arXiv",
    doi = "10.1103/PhysRevD.62.084003",
    journal = "Phys. Rev. D",
    volume = "62",
    pages = "084003",
    year = "2000"
}

@article{Capozziello:2024ucm,
    author = "Capozziello, Salvatore and De Bianchi, Silvia and Battista, Emmanuele",
    title = "{Avoiding singularities in Lorentzian-Euclidean black holes: The role of~atemporality}",
    eprint = "2404.17267",
    archivePrefix = "arXiv",
    primaryClass = "gr-qc",
    doi = "10.1103/PhysRevD.109.104060",
    journal = "Phys. Rev. D",
    volume = "109",
    number = "10",
    pages = "104060",
    year = "2024"
}

@article{Capozziello:2025wwl,
    author = "Capozziello, Salvatore and Battista, Emmanuele and De Bianchi, Silvia",
    title = "{Null geodesics, causal structure, and matter accretion in Lorentzian-Euclidean black holes}",
    eprint = "2507.08431",
    archivePrefix = "arXiv",
    primaryClass = "gr-qc",
    doi = "10.1103/ybjp-8w2w",
    journal = "Phys. Rev. D",
    volume = "112",
    number = "4",
    pages = "044009",
    year = "2025"
}

@article{DeBianchi:2025bgn,
    author = "De Bianchi, Silvia and Capozziello, Salvatore and Battista, Emmanuele",
    title = "{Atemporality from Conservation Laws of Physics in Lorentzian-Euclidean Black Holes}",
    eprint = "2504.17570",
    archivePrefix = "arXiv",
    primaryClass = "gr-qc",
    doi = "10.1007/s10701-025-00848-z",
    journal = "Found. Phys.",
    volume = "55",
    number = "3",
    pages = "36",
    year = "2025"
}

@article{Bozza:2002zj,
    author = "Bozza, V.",
    title = "{Gravitational lensing in the strong field limit}",
    eprint = "gr-qc/0208075",
    archivePrefix = "arXiv",
    doi = "10.1103/PhysRevD.66.103001",
    journal = "Phys. Rev. D",
    volume = "66",
    pages = "103001",
    year = "2002"
}

@article{Bozza:2010xqn,
    author = "Bozza, Valerio",
    title = "{Gravitational Lensing by Black Holes}",
    eprint = "0911.2187",
    archivePrefix = "arXiv",
    primaryClass = "gr-qc",
    doi = "10.1007/s10714-010-0988-2",
    journal = "Gen. Rel. Grav.",
    volume = "42",
    pages = "2269--2300",
    year = "2010"
}

@article{Ishihara:2016vdc,
    author = "Ishihara, Asahi and Suzuki, Yusuke and Ono, Toshiaki and Kitamura, Takao and Asada, Hideki",
    title = "{Gravitational bending angle of light for finite distance and the Gauss-Bonnet theorem}",
    eprint = "1604.08308",
    archivePrefix = "arXiv",
    primaryClass = "gr-qc",
    doi = "10.1103/PhysRevD.94.084015",
    journal = "Phys. Rev. D",
    volume = "94",
    number = "8",
    pages = "084015",
    year = "2016"
}

@article{Werner:2012rc,
    author = "Werner, M. C.",
    title = "{Gravitational lensing in the Kerr-Randers optical geometry}",
    eprint = "1205.3876",
    archivePrefix = "arXiv",
    primaryClass = "gr-qc",
    doi = "10.1007/s10714-012-1458-9",
    journal = "Gen. Rel. Grav.",
    volume = "44",
    pages = "3047--3057",
    year = "2012"
}

@article{Kumar:2020hgm,
    author = "Kumar, Rahul and Ghosh, Sushant G. and Wang, Anzhong",
    title = "{Gravitational deflection of light and shadow cast by rotating Kalb-Ramond black holes}",
    eprint = "2001.00460",
    archivePrefix = "arXiv",
    primaryClass = "gr-qc",
    doi = "10.1103/PhysRevD.101.104001",
    journal = "Phys. Rev. D",
    volume = "101",
    number = "10",
    pages = "104001",
    year = "2020"
}

@article{Bisnovatyi-Kogan:2010flt,
    author = "Bisnovatyi-Kogan, G. S. and Tsupko, O. Yu.",
    title = "{Gravitational lensing in a non-uniform plasma}",
    eprint = "1006.2321",
    archivePrefix = "arXiv",
    primaryClass = "astro-ph.CO",
    doi = "10.1111/j.1365-2966.2010.16290.x",
    journal = "Mon. Not. Roy. Astron. Soc.",
    volume = "404",
    pages = "1790--1800",
    year = "2010"
}

@article{Ono:2017pie,
    author = "Ono, Toshiaki and Ishihara, Asahi and Asada, Hideki",
    title = "{Gravitomagnetic bending angle of light with finite-distance corrections in stationary axisymmetric spacetimes}",
    eprint = "1704.05615",
    archivePrefix = "arXiv",
    primaryClass = "gr-qc",
    doi = "10.1103/PhysRevD.96.104037",
    journal = "Phys. Rev. D",
    volume = "96",
    number = "10",
    pages = "104037",
    year = "2017"
}

@article{Kuang:2022ojj,
    author = "Kuang, Xiao-Mei and Tang, Zi-Yu and Wang, Bin and Wang, Anzhong",
    title = "{Constraining a modified gravity theory in strong gravitational lensing and black hole shadow observations}",
    eprint = "2206.05878",
    archivePrefix = "arXiv",
    primaryClass = "gr-qc",
    doi = "10.1103/PhysRevD.106.064012",
    journal = "Phys. Rev. D",
    volume = "106",
    number = "6",
    pages = "064012",
    year = "2022"
}

@article{Nascimento:2020ime,
    author = "Nascimento, J. R. and Petrov, A. Yu. and Porfirio, P. J. and Soares, A. R.",
    title = "{Gravitational lensing in black-bounce spacetimes}",
    eprint = "2005.13096",
    archivePrefix = "arXiv",
    primaryClass = "gr-qc",
    doi = "10.1103/PhysRevD.102.044021",
    journal = "Phys. Rev. D",
    volume = "102",
    number = "4",
    pages = "044021",
    year = "2020"
}

@article{Fu:2021akc,
    author = "Fu, Qi-Ming and Zhao, Li and Liu, Yu-Xiao",
    title = "{Weak deflection angle by electrically and magnetically charged black holes from nonlinear electrodynamics}",
    eprint = "2101.08409",
    archivePrefix = "arXiv",
    primaryClass = "gr-qc",
    doi = "10.1103/PhysRevD.104.024033",
    journal = "Phys. Rev. D",
    volume = "104",
    number = "2",
    pages = "024033",
    year = "2021"
}

@book{Wald:1984rg,
    author = "Wald, Robert M.",
    title = "{General Relativity}",
    doi = "10.7208/chicago/9780226870373.001.0001",
    publisher = "Chicago Univ. Pr.",
    address = "Chicago, USA",
    year = "1984"
}

@article{Li:2019mqw,
    author = "Li, Zonghai and Zhou, Tao",
    title = "{Equivalence of Gibbons-Werner method to geodesics method in the study of gravitational lensing}",
    eprint = "1908.05592",
    archivePrefix = "arXiv",
    primaryClass = "gr-qc",
    doi = "10.1103/PhysRevD.101.044043",
    journal = "Phys. Rev. D",
    volume = "101",
    number = "4",
    pages = "044043",
    year = "2020"
}

@article{Rahaman:2010xs,
    author = "Rahaman, F. and Nandi, K. K. and Bhadra, A. and Kalam, M. and Chakraborty, K.",
    title = "{Perfect Fluid Dark Matter}",
    eprint = "1009.3572",
    archivePrefix = "arXiv",
    primaryClass = "gr-qc",
    doi = "10.1016/j.physletb.2010.09.038",
    journal = "Phys. Lett. B",
    volume = "694",
    pages = "10--15",
    year = "2011"
}

@article{Xu:2017bpz,
    author = "Xu, Zhaoyi and Wang, Jiancheng and Hou, Xian",
    title = "{Kerr{\textendash}anti-de Sitter/de Sitter black hole in perfect fluid dark matter background}",
    eprint = "1711.04538",
    archivePrefix = "arXiv",
    primaryClass = "gr-qc",
    doi = "10.1088/1361-6382/aabcb6",
    journal = "Class. Quant. Grav.",
    volume = "35",
    number = "11",
    pages = "115003",
    year = "2018"
}

@article{Hou:2018avu,
    author = "Hou, Xian and Xu, Zhaoyi and Wang, Jiancheng",
    title = "{Rotating Black Hole Shadow in Perfect Fluid Dark Matter}",
    eprint = "1810.06381",
    archivePrefix = "arXiv",
    primaryClass = "gr-qc",
    doi = "10.1088/1475-7516/2018/12/040",
    journal = "JCAP",
    volume = "12",
    pages = "040",
    year = "2018"
}

@article{Hou:2018bar,
    author = "Hou, Xian and Xu, Zhaoyi and Zhou, Ming and Wang, Jiancheng",
    title = "{Black hole shadow of Sgr A$^{*}$ in dark matter halo}",
    eprint = "1804.08110",
    archivePrefix = "arXiv",
    primaryClass = "gr-qc",
    doi = "10.1088/1475-7516/2018/07/015",
    journal = "JCAP",
    volume = "07",
    pages = "015",
    year = "2018"
}

@article{Haroon:2018ryd,
    author = "Haroon, Sumarna and Jamil, Mubasher and Jusufi, Kimet and Lin, Kai and Mann, Robert B.",
    title = "{Shadow and Deflection Angle of Rotating Black Holes in Perfect Fluid Dark Matter with a Cosmological Constant}",
    eprint = "1810.04103",
    archivePrefix = "arXiv",
    primaryClass = "gr-qc",
    doi = "10.1103/PhysRevD.99.044015",
    journal = "Phys. Rev. D",
    volume = "99",
    number = "4",
    pages = "044015",
    year = "2019"
}

@article{Gao:2023ltr,
    author = "Gao, Xiao-Jun and Yan, Xiao-kun and Yin, Yihao and Hu, Ya-Peng",
    title = "{Gravitational lensing by a charged spherically symmetric black hole immersed in thin dark matter}",
    eprint = "2303.00190",
    archivePrefix = "arXiv",
    primaryClass = "gr-qc",
    doi = "10.1140/epjc/s10052-023-11414-0",
    journal = "Eur. Phys. J. C",
    volume = "83",
    number = "4",
    pages = "281",
    year = "2023"
}

@article{Khriplovich:2008ij,
    author = "Khriplovich, I. B. and Pomeransky, A. A.",
    title = "{Does Cosmological Term Influence Gravitational Lensing?}",
    eprint = "0801.1764",
    archivePrefix = "arXiv",
    primaryClass = "gr-qc",
    doi = "10.1142/S0218271808013832",
    journal = "Int. J. Mod. Phys. D",
    volume = "17",
    pages = "2255--2259",
    year = "2008"
}

@article{Park:2008ih,
    author = "Park, Minjoon",
    title = "{Rigorous Approach to the Gravitational Lensing}",
    eprint = "0804.4331",
    archivePrefix = "arXiv",
    primaryClass = "astro-ph",
    doi = "10.1103/PhysRevD.78.023014",
    journal = "Phys. Rev. D",
    volume = "78",
    pages = "023014",
    year = "2008"
}

@article{Sereno:2007rm,
    author = "Sereno, M.",
    title = "{On the influence of the cosmological constant on gravitational lensing in small systems}",
    eprint = "0711.1802",
    archivePrefix = "arXiv",
    primaryClass = "astro-ph",
    doi = "10.1103/PhysRevD.77.043004",
    journal = "Phys. Rev. D",
    volume = "77",
    pages = "043004",
    year = "2008"
}

@article{Arakida:2011ty,
    author = "Arakida, Hideyoshi and Kasai, Masumi",
    title = "{Effect of the cosmological constant on the bending of light and the cosmological lens equation}",
    eprint = "1110.6735",
    archivePrefix = "arXiv",
    primaryClass = "gr-qc",
    doi = "10.1103/PhysRevD.85.023006",
    journal = "Phys. Rev. D",
    volume = "85",
    pages = "023006",
    year = "2012"
}

@article{Ishak:2008zc,
    author = "Ishak, Mustapha and Rindler, Wolfgang and Dossett, Jason",
    title = "{More on Lensing by a Cosmological Constant}",
    eprint = "0810.4956",
    archivePrefix = "arXiv",
    primaryClass = "astro-ph",
    doi = "10.1111/j.1365-2966.2010.16261.x",
    journal = "Mon. Not. Roy. Astron. Soc.",
    volume = "403",
    pages = "2152--2156",
    year = "2010"
}

@article{Bhadra:2010jr,
    author = "Bhadra, Arunava and Biswas, Swarnadeep and Sarkar, Kabita",
    title = "{Gravitational deflection of light in the Schwarzschild -de Sitter space time}",
    eprint = "1007.3715",
    archivePrefix = "arXiv",
    primaryClass = "gr-qc",
    doi = "10.1103/PhysRevD.82.063003",
    journal = "Phys. Rev. D",
    volume = "82",
    pages = "063003",
    year = "2010"
}

@article{Sereno:2008kk,
    author = "Sereno, M.",
    title = "{The role of Lambda in the cosmological lens equation}",
    eprint = "0807.5123",
    archivePrefix = "arXiv",
    primaryClass = "astro-ph",
    doi = "10.1103/PhysRevLett.102.021301",
    journal = "Phys. Rev. Lett.",
    volume = "102",
    pages = "021301",
    year = "2009"
}

@article{Ginsparg:1982rs,
    author = "Ginsparg, Paul H. and Perry, Malcolm J.",
    title = "{Semiclassical Perdurance of de Sitter Space}",
    reportNumber = "HUTP-82/A035",
    doi = "10.1016/0550-3213(83)90636-3",
    journal = "Nucl. Phys. B",
    volume = "222",
    pages = "245--268",
    year = "1983"
}

@article{Israel:1966rt,
    author = "Israel, W.",
    title = "{Singular hypersurfaces and thin shells in general relativity}",
    doi = "10.1007/BF02710419",
    journal = "Nuovo Cim. B",
    volume = "44S10",
    pages = "1",
    year = "1966",
    note = "[Erratum: Nuovo Cim.B 48, 463 (1967)]"
}

@misc{Ovgun:2025mdg,
    author = {{\"O}vg{\"u}n, Ali and Pantig, Reggie C.},
    title = "{Finite Distance Corrections to Vacuum Birefringence in Strong Gravitational and Electromagnetic Fields}",
    eprint = "2512.18727",
    archivePrefix = "arXiv",
    primaryClass = "gr-qc",
    month = "12",
    year = "2025"
}

@article{Lu:2025mcm,
    author = "Lu, Yi and Pan, Xiao-Yin and Lai, Meng-Yun and Wang, Qing-hai",
    title = "{Finite-distance gravitational lensing of a global monopole in a Schwarzschild{\textendash}de Sitter spacetime}",
    eprint = "2504.00777",
    archivePrefix = "arXiv",
    primaryClass = "gr-qc",
    doi = "10.1103/spt9-mpg2",
    journal = "Phys. Rev. D",
    volume = "112",
    number = "6",
    pages = "064051",
    year = "2025"
}

@article{Ahmed:2025ojg,
    author = "Ahmed, Faizuddin and Bouzenada, Abdelmalek and Silva, Edilberto O.",
    title = "{AdS black hole solution with a dark matter halo surrounded by a cloud of strings}",
    eprint = "2509.10829",
    archivePrefix = "arXiv",
    primaryClass = "gr-qc",
    doi = "10.1140/epjc/s10052-025-15113-w",
    journal = "Eur. Phys. J. C",
    volume = "85",
    number = "12",
    pages = "1385",
    year = "2025"
}

@article{Sarkar:2025fao,
    author = "Sarkar, Susmita and Sarkar, Nayan and Shah, Hasrat Hussian and Balo, Pankaj and Rahaman, Farook",
    title = "{Deflection angle of regular black holes in nonlinear electrodynamics: Gauss-Bonnet theorem, time delay, shadow, and greybody bound}",
    eprint = "2509.02633",
    archivePrefix = "arXiv",
    primaryClass = "gr-qc",
    doi = "10.1016/j.physletb.2025.139905",
    journal = "Phys. Lett. B",
    volume = "870",
    pages = "139905",
    year = "2025"
}

\end{document}